\documentclass[article,reprint,amsmath,amssymb,superscriptaddress,longbibliography]{revtex4-1}
\usepackage[colorlinks,
linkcolor=blue,
anchorcolor=blue,
citecolor=blue,
]{hyperref}

\usepackage{graphicx}% Include figure files untitled
\usepackage{epstopdf}
\usepackage{hyperref}
\usepackage{upgreek}
\usepackage{breakurl}
\usepackage{multirow}
\usepackage{enumitem}

\usepackage{tablefootnote}
\usepackage{threeparttable}
\usepackage{tabularx}
\usepackage{booktabs}
\usepackage{textcomp}
\usepackage{lineno}%\linenumbers
\setcitestyle{super}

\begin{document}

%\title {A doping threshold for polar metals}
%\title {Dipolar-Kondo and RKKY interaction across polar to non-polar transition in a polar metal}
\title {Clues to potential dipolar-Kondo and RKKY interactions in a polar metal}

\author{Xiaohui Yang}
%\thanks{Equal contributions}
%\email{yangxiaohui@westlake.edu.cn}
 \affiliation{Key Laboratory for Quantum Materials of Zhejiang Province, Department of Physics, School of Science, Westlake University, Hangzhou 310030, P. R. China}
 \affiliation{Institute of Natural Sciences, Westlake Institute for Advanced Study, Hangzhou 310024, P. R. China}
 \affiliation{Department of Physics, China Jiliang University, Hangzhou 310018, Zhejiang, P. R. China}

\author{Wanghua Hu}
%\thanks{Equal contributions}
 \affiliation{Key Laboratory for Quantum Materials of Zhejiang Province, Department of Physics, School of Science, Westlake University, Hangzhou 310030, P. R. China}
 \affiliation{Institute of Natural Sciences, Westlake Institute for Advanced Study, Hangzhou 310024, P. R. China}
 \affiliation{Department of Physics, Fudan University, Shanghai 200433, P. R. China}

\author{Jialu Wang}
 \affiliation{Key Laboratory for Quantum Materials of Zhejiang Province, Department of Physics, School of Science, Westlake University, Hangzhou 310030, P. R. China}
 \affiliation{Institute of Natural Sciences, Westlake Institute for Advanced Study, Hangzhou 310024, P. R. China}

\author{Zhuokai Xu}
 \affiliation{Key Laboratory for Quantum Materials of Zhejiang Province, Department of Physics, School of Science, Westlake University, Hangzhou 310030, P. R. China}
 \affiliation{Institute of Natural Sciences, Westlake Institute for Advanced Study, Hangzhou 310024, P. R. China}

\author{Tao Wang}
 \affiliation{Key Laboratory for Quantum Materials of Zhejiang Province, Department of Physics, School of Science, Westlake University, Hangzhou 310030, P. R. China}
 \affiliation{Institute of Natural Sciences, Westlake Institute for Advanced Study, Hangzhou 310024, P. R. China}
 \affiliation{Department of Physics, Fudan University, Shanghai 200433, P. R. China}

\author{Zhefeng Lou}
\affiliation{Key Laboratory for Quantum Materials of Zhejiang Province, Department of Physics, School of Science, Westlake University, Hangzhou 310030, P. R. China}
\affiliation{Institute of Natural Sciences, Westlake Institute for Advanced Study, Hangzhou 310024, P. R. China}

\author{Xiao Lin}
 \email{linxiao@westlake.edu.cn}
 \affiliation{Key Laboratory for Quantum Materials of Zhejiang Province, Department of Physics, School of Science, Westlake University, Hangzhou 310030, P. R. China}
 \affiliation{Institute of Natural Sciences, Westlake Institute for Advanced Study, Hangzhou 310024, P. R. China}

\date{\today}

\begin{abstract}
\noindent
The coexistence of electric dipoles and itinerant electrons in a solid was postulated decades ago, before being experimentally established in several `polar metals' during the last decade. Here, we report a concentration-driven polar-to-nonpolar phase transition in electron-doped BaTiO$_3$. Comparing our case with other polar metals, we find a particular threshold concentration ($n^*$) linked to the dipole density ($n_\textrm{d}$). The universal ratio $\frac{n_\textrm{d}}{n^*}\approx 8.0(6)$ suggests a common mechanism across different polar systems, possibly explained by a dipolar Ruderman-Kittel-Kasuya-Yosida theory. 
Moreover, in BaTiO$_3$, we observe enhanced thermopower and upturn on resistivity at low temperatures near $n^*$, resembling the Kondo effect. We argue that local electric dipoles act as two-level-systems, whose fluctuations couple with surrounding electron clouds, giving rise to a potential dipolar-counterpart of the Kondo effect. Our findings unveil a mostly uncharted territory for exploring emerging physics associated with electron-dipole correlations, encouraging further theoretical work on dipolar-RKKY and Kondo interactions.

\end{abstract}

\maketitle

\noindent\textbf{INTRODUCTION}\\
\noindent In 1965, Anderson and Blount proposed a concept of a `ferroelectric (FE)' or polar metal~\cite{Anderson1965PRL}, for which a structural phase transition breaks the inversion symmetry, resulting in a polar axis in a metal. This envisage was overlooked by decades, owing to the common knowledge that mobile electrons provide strong screening for Coulomb interactions. Recently, in LiOsO$_3$, a stoichiometric metal, convergent-beam electron diffraction detected a phase transition, structurally indistinguishable from the FE transition of its insulating cousins, LiNbO$_3$ and LiTaO$_3$~\cite{ShiYG2013NM}. The existence of polar order could be interpreted by decoupling between itinerant electrons and soft transverse optical phonons~\cite{Puggioni2014NC,Laurita2019NC}. It is now known that polar metals have been observed in a handful of materials with explicit experimental evidences~\cite{Cobden2018Nature,Cai2021NC,Rhodes2023nature}, including 2D WTe$_2$~\cite{Cobden2018Nature}, Hg$_3$Te$_2$X$_2$ (X= Cl, Br)~\cite{Cai2021NC}, etc and dozens of polar metal candidates were summarized in recent papers~\cite{RondinelliPRM2023,LiuS2021}

There is another group of polar metals stemming from charge doping in ferroelectrics, also dubbed as extrinsic polar metals or degenerately doped ferroelectrics~\cite{RondinelliPRM2023}. For instance, in BaTiO$_3$ (BTO)~\cite{Kolodiazhnyi2010PRL} and Sr$_\textrm{1-x}$Ca$_\textrm{x}$TiO$_3$ (SCTO)~\cite{Rischau2017NP}, electron doping renders the FE insulator a dilute metal, in which the Fermi sea is too shallow to fully screen the long range polar order. Of particular interest is the ability to tuning the polar phase in these systems. In SCTO, the polar transition is continuously suppressed to zero temperature ($T$) as the electron concentration ($n$) approaches to a doping threshold ($n^*$). Amazingly, near $n^*$,  the superconductivity (SC) in SCTO is enhanced in comparison with that in its non-polar counterpart SrTiO$_3$ (STO)~\cite{Rischau2017NP}, implying the link between SC and polar threshold~\cite{Rischau2017NP,Marel2016SR,Balatsky2015PRL,Ruhman2019PRX}.

The study of polar metals remains at the early stage. Little is known about the interplay of electric dipoles in a Fermi sea. How does it influence the polar transition? Is there a common origin underlying the polar to non-polar (P-NP) transition in different polar metals? Is this interplay a many-body effect? If yes, will it lead to emerging physics associated with electron correlations?

\begin{figure*}[hpbt]
	\begin{center}
		%	\hspace{-0.5 cm}
		\includegraphics[width=17.3cm]{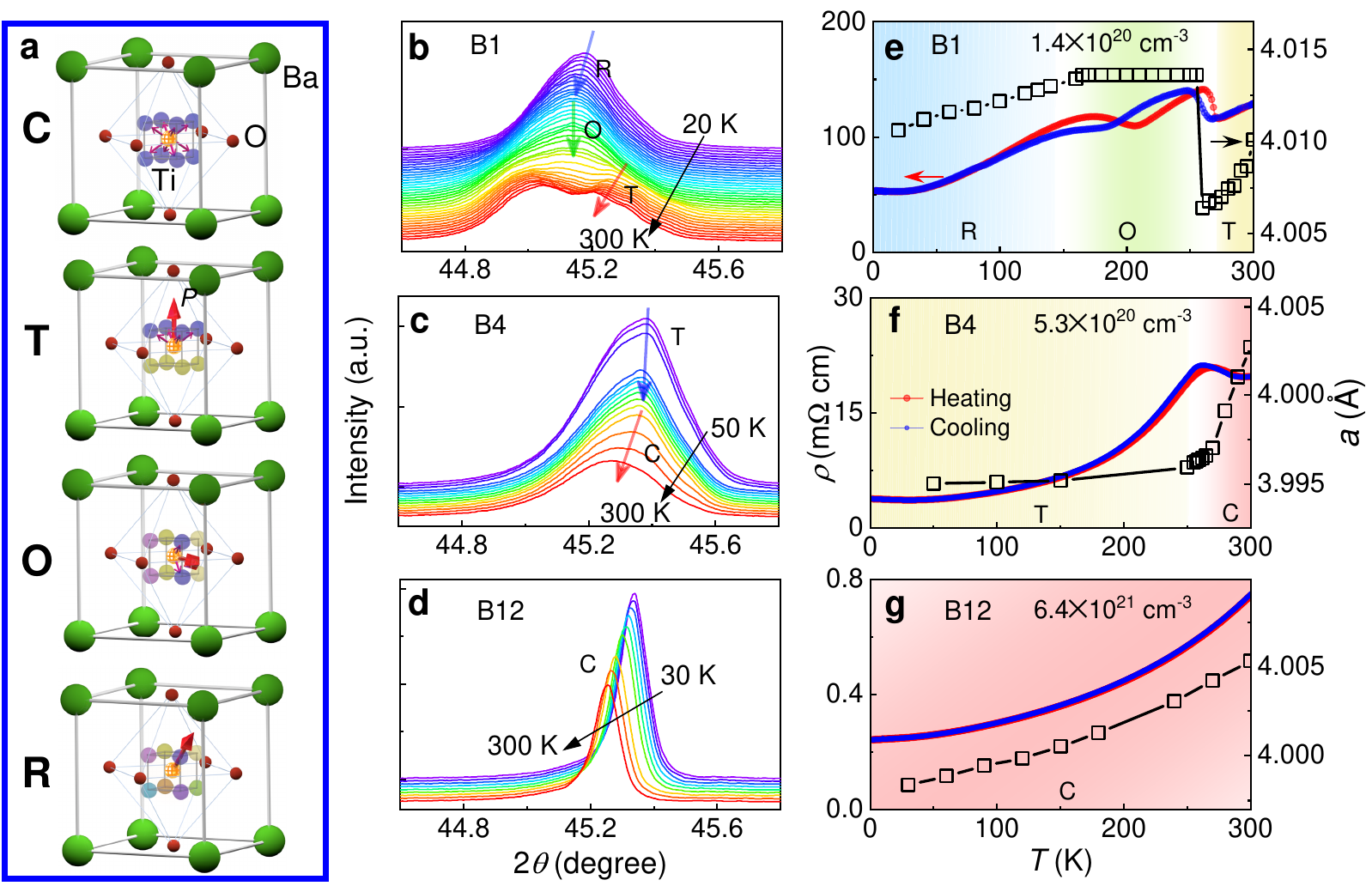}
	\end{center}
	\setlength{\abovecaptionskip}{-8 pt}
	\caption{\textbf{Evolution of polar transitions in BTO$_{3-\delta}$.} \textbf{a} Configuration of Ti ions in \textbf{C}, \textbf{T}, \textbf{O}, \textbf{R} phases. For \textbf{C} phase, Ti occupies eight equivalent off-center positions, thus without polarization ($P$); \textbf{T} phase: two sets of four equivalent positions with out-of-plane $P$; \textbf{O} phase: four sets of two equivalent positions with $P$ along [110] direction; \textbf{R} phase: eight inequivalent positions with $P$ along [111] direction. \textbf{b-d} Enlarged figure of (200) XRD peaks for B1, B4 and B12 at various $T$. The color arrows guide polar transitions. XRD were performed on heating. \textbf{e-g} Extracted lattice constant ($a$) versus $T$, in comparison with resistivity ($\rho$). $\rho$ was measured both on cooling and heating. %(b) Relative permittivity ($\varepsilon_\textrm{r}$) versus $T$ for an insulating specimen. The insets shows hysteresis at two transitions. The dashed line is from Ref \cite{Akishige1998FE}, which highlights the high quality of our specimen.
	}
	\label{Fig1}
\end{figure*}

To address these questions, we present a study of P-NP transition in BaTiO$_{3-\delta}$ (BTO$_{3-\delta}$), in comparison with previous reports on Sr$_\textrm{1-x}$Ca$_\textrm{x}$TiO$_{3-\delta}$ (SCTO$_{3-\delta}$)~\cite{WangJL2019npj} and Sr$_\textrm{1-x}$Ca$_\textrm{x}$Ti$_\textrm{1-y}$Nb$_\textrm{y}$O$_3$ (SCTNO)~\cite{Inoue2022npj}. We find the polar phase terminates at a threshold doping, when one electron is met by about eight dipoles. This linearity between the dipole density ($n_\textrm{d}$) and $n^*$ is robust in spite of distinct polar nature in three systems. Such a remarkable experimental evidence points to a common cause for P-NP transitions in polar metals. The ratio $\frac{n_\textrm{d}}{n^*}\approx 8.0(6)$ appears to correspond to destructive interference of Friedel oscillations~\cite{Friedel1958} generated by neighboring dipoles inside a Fermi sea via the dipolar counterpart of Ruderman-Kittel-Kasuya-Yosida (RKKY) interaction~\cite{Glinchuk1992RKKY}. Near $n^*$, anomalously enhanced thermoelectric response ($S$) and upturn on resistivity ($\rho$) at low-$T$ are observed in BTO$_{3-\delta}$. 
\textcolor{black}{This intriguing observation strongly suggests the manifestation of the Kondo effect. Given its proximity to the polar threshold, we put forward the possibility of a dipolar-Kondo phenomenon, a many-body resonant scattering stemming from the strong coupling between dipole fluctuations and Friedel oscillations. A comprehensive theoretical elucidation of this concept is yet to explored.}

%Our findings opens up a new venue for the study of correlation physics.

\vspace{3ex}
\noindent\textbf{RESULTS}\\
\noindent\textbf{Polar phase transitions in BTO}\\
BTO is one of the most studied lead-free FE oxides~\cite{Choi2004Science}. It is a cubic perovskite at $T>403$ K~\cite{Itoh1985FE} and undergoes a cascade of structural phase transitions from paraelectric cubic phase (\textbf{C}, $Pm\overline{3}m$) to FE tetragonal (\textbf{T}, $P4mm$), orthorhombic (\textbf{O}, $Amm2$), and rhombohedral (\textbf{R}, $R3m$) phases as $T$ lowers~\cite{Buttner1992Acta} (see Supplementary Note~1 and Supplementary Fig.~1 for dielectric measurements). In Fig.~\ref{Fig1}a, FE transitions involve the confinement of Ti ion dynamic hopping on eight-fold off-center positions, which displays dual feature of order-disorder and displacive ferroelectricity ~\cite{Pasciak2018PRL,Hlinka2008PRL}.

Upon electron doping, the polar phase in BTO$_{3-\delta}$ is robust with $n$ up to $3\times10^{20}~\textrm{cm}^{-3}$~\cite{Kolodiazhnyi2010PRL}. Empirically, BTO$_{3-\delta}$ transforms from perovskite to hexagonal phase at sightly higher $n$~\cite{Kolodiazhnyi2008PRB,Sinclair1999JMC}. By improving annealing methods, we prepared twelve BTO$_{3-\delta}$ perovskite single crystals (B1-B12) with $n$ more than one order of magnitude higher (see Supplementary Notes~2,3, Supplementary Figs.~2,3 and Supplementary Table~1 for details), which endows us an opportunity to study the P-NP transition.

\begin{figure*}[hpbt]
	\begin{center}
		\includegraphics[width=17.3cm]{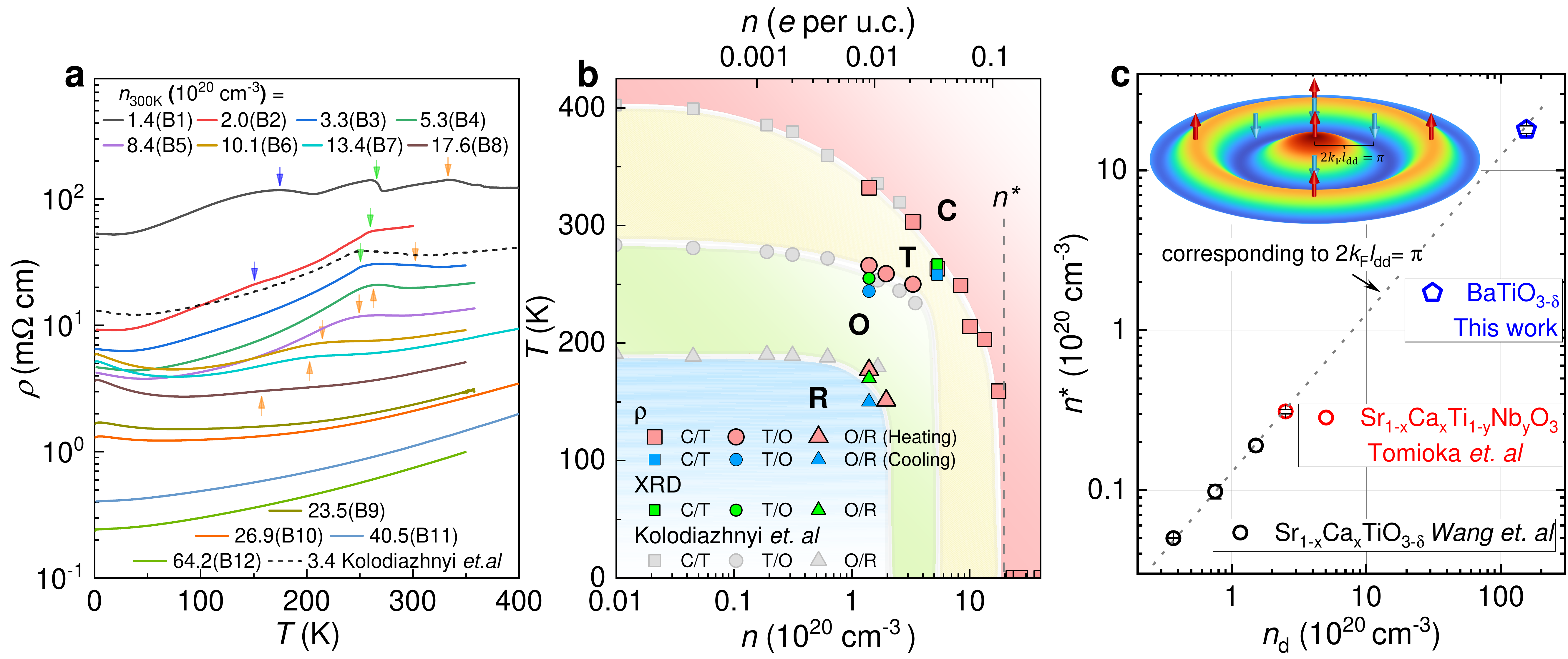}
	\end{center}
	\setlength{\abovecaptionskip}{-8 pt}
	\caption{\textbf{Polar phase diagram.} \textbf{a} $\rho$ versus $T$ for all specimens from 2 K to 400 K. The orange, green, blue arrows mark \textbf{C}/\textbf{T}, \textbf{T}/\textbf{O} and \textbf{O}/\textbf{R} transitions, respectively. The dotted curve is from Ref.~\citenum{Kolodiazhnyi2010PRL}. \textbf{b} Polar phase diagram for BTO$_{3-\delta}$. The data are collected from XRD and resistivity measurements. The grey points are referred from Ref.~\citenum{Kolodiazhnyi2010PRL}. The vertical dashed line marks $n^*$.  \textbf{c} $n^*$ scales with $n_\textrm{d}$. Three black circles were collected from Sr$_\textrm{1-x}$Ca$_\textrm{x}$TiO$_3$ with $x=0.22\%, 0.45\%$ and $0.9\%$ in a previous report~\cite{WangJL2019npj}. The red circle was collected from SCTNO~\cite{Inoue2022npj}. For SCTO and SCTNO, $n_\textrm{d}=x/a^3_\textrm{STO}$, $a_\textrm{STO}=3.905~\textrm{\AA}$ is the lattice constant of STO. For BTO, $n_\textrm{d}=1/a^3_\textrm{BTO}$, $a_\textrm{BTO}=4.009\textrm{\AA}$. The dotted line is a linear fit. The inset is an illustration depicting the dipolar-RKKY interaction. Anti-FE and FE couplings alternate radically in real space. The error bars represent the uncertainty in determining the threshold concentration.
	}
	\label{Fig2}
\end{figure*}

$In~situ$ variable-$T$ X-ray diffraction (XRD) at (200) peak is shown in Fig.~\ref{Fig1}b-d for three representatives (B1, B4 and B12). See the full range data in Supplementary Fig.~4. In Fig.~\ref{Fig1}b, for B1 with $n\approx1.4\times 10^{20}$ cm$^{-3}$, the arrows mark \textbf{T}/\textbf{O} and \textbf{O}/\textbf{R} transitions at about 255~K and 170~K, respectively, as expected in Ref.~\citenum{Kolodiazhnyi2010PRL}. At high-$T$, two split peaks appear owing to multidomain structures, that is common in \textbf{T}-phase~\cite{Jung2002MRB}. The extracted lattice constant ($a$) versus $T$ is presented in Fig.~\ref{Fig1}e. $a$ shows a discontinuous jump at \textbf{T}/\textbf{O} and slope change at \textbf{O}/\textbf{R} transition. The transitions are also monitored by $\rho$. Two kinks with apparent thermal hysteresis are observed at $T_\textrm{\textbf{T}/\textbf{O}}$ and $T_\textrm{\textbf{O}/\textbf{R}}$, implying first-order phase transitions.

XRD patterns of B4 are presented in Fig.~\ref{Fig1}c, with $n\approx5.3\times 10^{20}$ cm$^{-3}$ beyond the range of previous reports~\cite{Kolodiazhnyi2010PRL,Kolodiazhnyi2008PRB}. In Fig.~\ref{Fig1}f, $a$ decreases smoothly as $T$ lowers and changes slope at 267~K, corresponding to \textbf{C}/\textbf{T} transition. It is manifested by an anomaly on $\rho$ with shrunk thermal hysteresis, implying a weak first-order transition. In Fig.~\ref{Fig1}d and \ref{Fig1}g,  $a$ and $\rho$ evolve smoothly for B12 with $n\approx6.4\times 10^{21}$ cm$^{-3}$, indicating the absence of polar phase.

\vspace{2ex}
\noindent\textbf{Doping threshold in polar phase diagram}\\
To obtain the polar phase diagram, $\rho$ versus $T$ for all twelve specimens is presented in Fig.~\ref{Fig2}a. For B1 with the lowest $n$, remarkable anomalies, marked by three arrows, are observed corresponding to \textbf{C}/\textbf{T}, \textbf{T}/\textbf{O} and \textbf{O}/\textbf{R} transitions. At higher $n$ (B4 - B8, $5.3 \times 10^{20}$ to $1.76 \times 10^{21}$ cm$^{-3}$), only a single anomaly is detected for \textbf{C}/\textbf{T} transition. With $n>2 \times 10^{21}$ cm$^{-3}$ (B9 - B12), there are no distinguishable anomalies, indicating the absence of polar transition. Note that the low-$T$ upturn is not a transition and will be discussed in detail later.

Figure~\ref{Fig2}b presents the phase diagram with the data extracted from Fig.~\ref{Fig2}a, which smoothly extends what was reported in Ref.~\citenum{Kolodiazhnyi2010PRL}. See Supplementary Note~4 and Supplementary Fig.~5 for the determination of critical temperature. As $n$ rises, polar transitions are suppressed to low-$T$ and terminate at $n^* \approx 1.8(1) \times 10^{21}$ cm$^{-3}$, corresponding to \textcolor{black}{0.116}$e$ per unit cell (per u.c.), that matches the predicted value ($n_\textrm{cal}$) of DFT calculations~\cite{WangY2012PRL}. One expected the polar phase is destroyed by strong Thomas-Fermi (TF) screening, given the calculated TF screening length ($r_\textrm{TF}\approx5~\textrm{\AA}$) comparable to $a$ (i.e. the inter-dipole distance -- $l_\textrm{dd}$)~\cite{Kolodiazhnyi2010PRL,WangY2012PRL}. However, we argue that this discussion overlooked the following two facts. First,  the TF screening is a special case of the Lindhard theory in a static and long-distance limit. The length scale of interest, here $l_\textrm{dd}$, should be much larger than the Fermi wavelength ($\lambda_\textrm{F}$) and then $a$. Accordingly, the scheme of TF screening in BTO should be concerned with caution (see more in Supplementary Note~5 and Supplementary Table~2.). Second, oxygen vacancies ($V_\textrm{O}$) are prevailing in present system. The impact of lattice defects on polar order might not be neglected~\cite{Iwazaki2012PRB}. If this is true, the consistency between $n^*$ and $n_\textrm{cal}$ seems like a coincidence. Subsequent DFT calculations suggested that electron doping disrupts the local off-centering Ti-O bonds, leading to the suppression of polar transitions~\cite{Tangney2021PRM,RondinelliPRB2020}.  Most recently, Gu \textit{et. al.}, proposed a mechanism involving the formation of polaronic quasi-particles made of the carriers and their surrounding dipoles~\cite{Gu2023Polaron}.

\begin{figure*}[hpbt]
	\begin{center}
		\vspace{-0.5 cm}
		%\flushleft{10.0 cm}
		%\hspace{-0.5cm}
		\includegraphics[width=15cm]{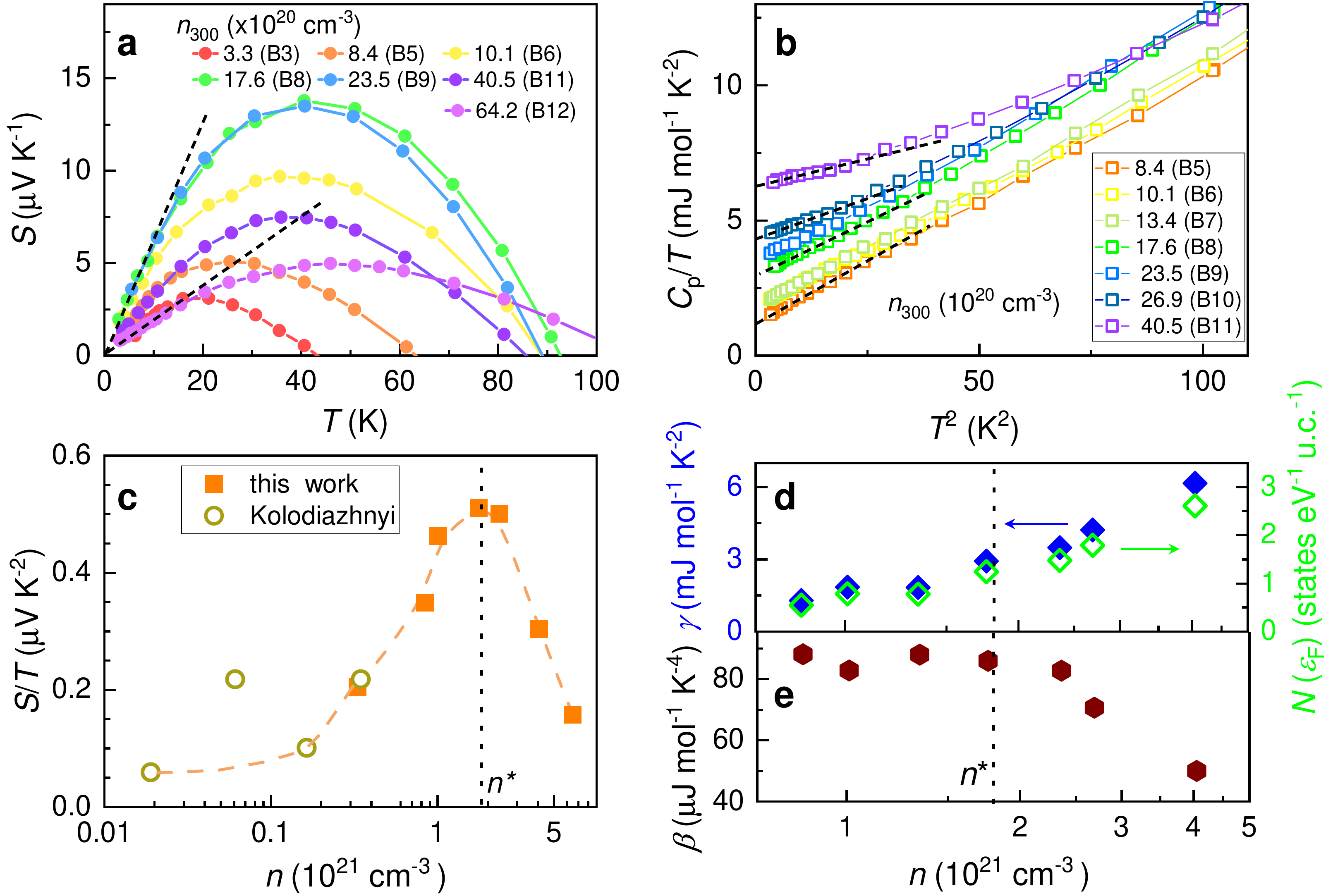}
	\end{center}
	\setlength{\abovecaptionskip}{-8 pt}
	\caption{\label{TEP} \textbf{Thermoelectric and thermodynamic properties.} \textbf{a} Seebeck coefficients ($S$) as a function of $T$ at low-$T$. The dashed lines mark the slope of $S$. \textbf{b} Slope of $S$ approaching to zero-$T$ as a function of $n$. The open points are referred to Ref.~\citenum{Kolodiazhnyi2008PRB}. \textbf{c} Heat capacity over $T$ ($C_\textrm{p}/T$) as a function of $T$-square. The dashed lines are linear fits. \textbf{d} $\gamma$ and the corresponding DOS ($N(\varepsilon_\textrm{F})$) as a function of $n$, for which $\gamma=\frac{\uppi^2 k_\textrm{B}^2}{3} N_\textrm{A} N(\varepsilon_\textrm{F})$, wherein $N_\textrm{A}$ is the Avogadro number and $k_\textrm{B}$ is the Boltzmann constant. \textbf{e} Phonon contribution ($\beta$) as a function of $n$.
	}
\end{figure*}

Below, we offer an alternative interpretation. In Fig.~\ref{Fig2}c, $n^*$ is plotted as a function of $n_\textrm{d}$ for BTO$_{3-\delta}$, which is compared with that of other polar metals: SCTO$_{3-\delta}$~\cite{Rischau2017NP,WangJL2019npj} and  SCTNO~\cite{Inoue2022npj}. $n^*$ is proportional to $n_\textrm{d}$, even if $n_\textrm{d}$ varies by almost three orders of magnitude.  Its slope indicates that the polar phase fades away when one electron is met by about 8.0(6) dipoles. The robust linearity with a ratio $\frac{n_\textrm{d}}{n^*}\approx \textcolor{black}{8.0(6)}$ across these systems is striking in view of the following distinct details. First, BTO is a strong FE with dense dipoles, correspondingly one dipole per u.c., composed of Ti off-center occupation with respect to the center of the unit cell. While, SCTO are weak FE and the dipoles are dilute, from Ca displacement off the corner Sr-sites  ($n_\textrm{d}=n_\textrm{Ca}$). Second, electron donors are different. For BTO and SCTO, it's $V_\textrm{O}$. While for SCTNO, it's Nb substitution of Ti ions. Thus, the linearity is insensitive to different kinds of lattice defects. These remarkable experimental facts highly suggest a common origin underlying the destruction of polar phase in polar metals. Here, we reach the first main outcome of the paper.

% \textcolor{red}{\sout{(i.e. one electron by two dipoles at each dimension)}}

Such a universality puts stringent constraints on relevant theories. TF screening fails to account for this linearity, since $r_\textrm{TF}$ scales with $n^{-1/6}(\varepsilon/m)^{1/2}$, wherein permittivity ($\varepsilon$) and effective electron mass ($m$) are material dependent. In 1992, Glinchuk \textit{el al.} proposed a \textcolor{black}{primitive} theory depicting the interplay of electric dipoles immersed a Fermi sea~\cite{Glinchuk1992RKKY} , via a dipolar analogue of RKKY interaction~\cite{RKKY1954}. This scheme was invoked by part of the authors attempting to explain the case of SCTO$_{3-\delta}$ ~\cite{Rischau2017NP,WangJL2019npj}. The interaction is expressed by
\begin{equation}
	V_\textrm{dd}\sim\frac{\textrm{cos} (2k_\textrm{F}l_\textrm{dd})}{l_\textrm{dd}^3}
\end{equation}
\noindent where $k_\textrm{F}$ is the Fermi wave vector and $l_\textrm{dd}=n_\textrm{d}^{-1/3}$.  Assuming an isotropic Fermi surface ($k_\textrm{F}=(3\uppi^2n)^{1/3}$),  $V_\textrm{dd}$ approaches to the first minimum at $n=n^*$ as seen in the inset of Fig.~\ref{Fig2}c, at which the anti-parallel alignment of the nearest dipoles is energetically favorable that inclines to interrupt the polar phase.

\begin{figure*}[hpbt]
	\begin{center}
		\vspace{-0.5 cm}
		%\flushleft{10.0 cm}
		%\hspace{-0.5cm}
		\includegraphics[width=17cm]{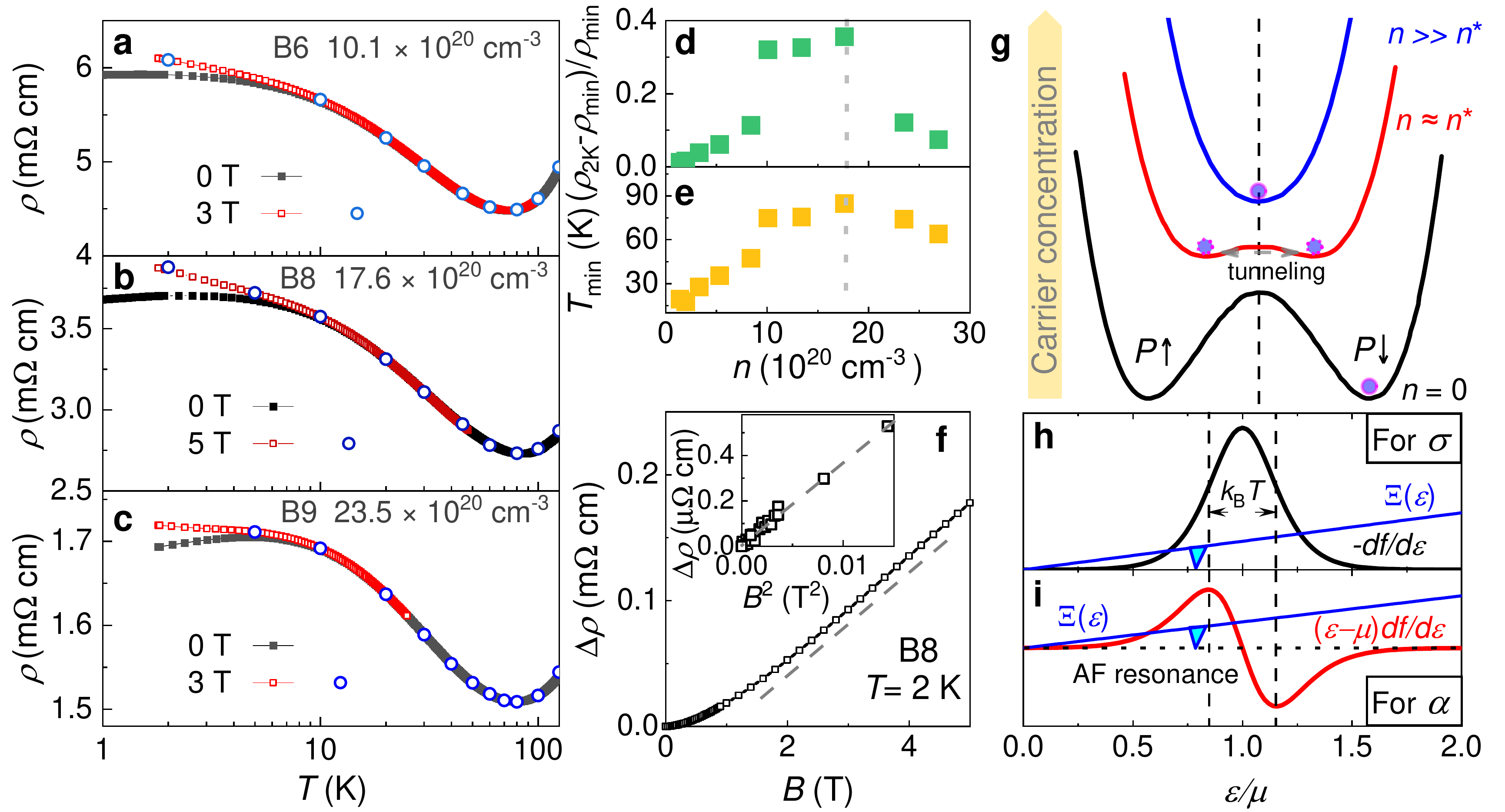}
	\end{center}
	\setlength{\abovecaptionskip}{-8 pt}
	\caption{\label{Fig4} \textbf{Low-$T$ upturn on resistivity.} \textbf{a-c} $\rho$ versus $T$ at $B=$ 0~T and 3/5~T below 125 K on a semi-log scale for B6, B8 and B9. The open circles are symmetrized data extracted from magnetoresistance (MR, see Supplementary Fig.~9).  \textbf{d-e} $(\rho_\textrm{2K}-\rho_\textrm{min})/\rho_\textrm{min}$ and $T_\textrm{min}$ as a function of $n$. \textbf{f} MR for B8 at 2~K. The inset is MR as a function of $B^2$ at $B\rightarrow0$. The dashed lines are guides to eyes. 
\textbf{g} Evolution of local potential with respect to Ti-ion displacement. \textbf{h-i} Kernels of Eq.~\ref{Eq2} defining transport coefficients. A Kondo effect in a TLS produces an Anderson-Friedel (AF) resonance near the Fermi level.
	}
\end{figure*}

\vspace{3ex}
\noindent\textbf{Enhancement of thermoelectric power}\\
Having established a universal doping threshold for P-NP transition, let us now explore exotic phenomena close to $n^*$ in BTO$_{3-\delta}$. In Fig.~\ref{TEP}a, we present $S$ as a function of $T$ at low-$T$ for different samples. $S$ evolves from negative to positive by decreasing $T$ (See the full-$T$ range in Supplementary Note~6 and Supplementary Fig.~6) and becomes asymptotically $T$-linear as $T$ approaches to zero. In Fig.~\ref{TEP}b, the slope of diffusive component ($S/T$) at $T \rightarrow 0$ K is extracted and presented versus $n$, in combining the data from Ref.~\citenum{Kolodiazhnyi2008PRB}. Surprisingly, $S/T$ enhances by one order of magnitude from $n\approx 10^{19}$ cm$^{-3}$ to the threshold region and then descends rapidly beyond. This nonmonotonic behavior is distinct from what one generally expected in Fermi liquids: as in STO, the diffusive $S/T$ decreases monotonically as $n$ rises~\cite{Lin2013PRX}. In a variety of correlated metals, $S/T$ in the zero-$T$ limit is inversely proportional to the Fermi energy ($\varepsilon_\textrm{F}$)~\cite{Behnia2004JPCM,Lin2013PRX}.  It's thus tantalizing to relate the enhancement of $S/T$ to re-normalization of $\varepsilon_\textrm{F}$ (i.e density of state -- DOS) owing to strong electron correlations. Critically strengthened correlations are characteristic of quantum criticality as in magnetic systems~~\cite{Lohneysen2007RMP}.

For more information, we present the heat capacity ($C_\textrm{p}$) in Fig.~\ref{TEP}c. A fit of $C_\textrm{p}/T$ to the formula  $C_\textrm{p}/T=\gamma+\beta T^2$ at $T\rightarrow0$~K deduces the Sommerfeld coefficients ($\gamma$) and the phonon term ($\beta$). In Fig.~\ref{TEP}d, $\gamma$ and the corresponding DOS evolve smoothly with the absence of singularity near $n^*$, strongly against quantum critical descriptions. In Fig.~\ref{TEP}e, $\beta$ is almost constant below $n\sim n^*$ and declines afterwards, that is unlikely a manifestation of structural quantum phase transition either, because the softening of boundary phonon modes will produce a peak on $\beta$ at the critical region~\cite{Goh2015PRL,Gruner2017NP}.

\vspace{2ex}
\noindent\textbf{Low-$T$ upturn on resistivity}\\
On the other hand, in Supplementary Fig.~7, the sign mismatch between the Hall coefficient ($R_\textrm{H}$) and $S$ at low-$T$ for B8 hints that there might be a resonance feature near the Fermi level~\cite{SunPJ2013PRL,Behniabook2015}. Thereby, we plot $\rho$ versus $T$ on a semi-log scale for $T< 125$ K in Fig.~\ref{Fig4}a-c for B6, B8 and B9. More data is included in Supplementary Fig.~8. $\rho$ exhibits a minimum ($\rho_\textrm{min}$) at a characteristic temperature ($T_\textrm{min}$). Below $T_\textrm{min}$, $\rho$ shows a distinct upturn, roughly following a $\textrm{log}T$ behavior and then levels off at lower-$T$ (Certain samples near $n^*$ show a downturn as $T\rightarrow0$~K, see discussion below). We take $(\rho_\textrm{2K}-\rho_\textrm{min})/\rho_\textrm{min}$ and $T_\textrm{min}$ as a crude measure of this effect. Similar to $S/T$, both terms maximize near $n^*$ in Fig.~\ref{Fig4}d-e, implying the intimacy between these phenomena and polar threshold.

This upturn could be attributed to a number of well-established mechanisms, such as the Kondo effect~\cite{Kondo1964spin}, weak localization (WL)~\cite{Lee1985RMP} and the Althshuler-Aronov (AA) effect~\cite{Altshuler1985book}. In Fig.~\ref{Fig4}a-c, the upturn remains intact under a magnetic field ($B=$ 3 or 5 T), except for an upward evolution below 10 K. In Fig.~\ref{Fig4}f, the unnegligible positive magnetoresistance (MR) below 10 K follows a $B^2$ dependence at $B\rightarrow0$ T and deviates at higher $B$ (becomes roughly linear) for B8, which is common in metals and presumably arises from orbital effects. These observations clearly rule out spin-Kondo and WL effects, since both mechanisms show substantial negative MR and suppress the upturn under such field. For the AA effect, the electron-electron interaction in disordered metals suppresses DOS at Fermi level ($\textrm{DOS}\sim\sqrt{|\varepsilon-\varepsilon_\textrm{F}|}$) and leads to a modification of $\rho$~\cite{Altshuler1985book}. It's unlikely that this effect could explain the anomalous enhancement of $S$ near $n^*$~\cite{Hsu1989PRB}. Furthermore, WL or AA effects would cause a $B^{1/2}$ dependence of MR~\cite{Altshuler1985book}, inconsistent with the observation.

Kondo physics exists in an arbitrary two-level-system (TLS)~\cite{VladarPRB1983,ZawadowskiAP1998}, non-exclusive to spins. Non-magnetic Kondo effect, e.g. orbital Kondo effect~\cite{JarilloNature2005} and charge Kondo effect~\cite{MatsushitaPRL2005}, has been explicitly documented in a variety of systems. In polar metals, it is intuitive to associate the TLS with dipoles. Figure~\ref{Fig4}g sketches the evolution of local potential surface with respect to Ti-ion displacement with $n$ in BTO. At $n=0$, there is a deep double-well with two minima equidistant from the lattice center. The Ti ion resides on one of the two minina, each minimum representing $P${\textuparrow} or $P${\textdownarrow}. This is naturally a TLS. As $n$ approaches to $n^*$, the large potential barrier is suppressed, resulting in a shallow double-well, which allows for quantum tunneling of Ti ions between two minima (dipole fluctuations), reminiscent of what is in quantum paraelectrics~\cite{Tangney2021PRM}. At $n\gg n^*$, the double-well merges into a single potential well, corresponding to a paraelectric phase.

A many-body resonant scattering involving the coupling between tunneling centers and the Fermi sea might give rise to a dipolar analogue of Kondo effect. Near $n^*$, this effect maximizes due to the critical enhanced number of tunneling centers, which is in line with our observations. Given the magnitude of upturn on resistivity $\Delta\rho=\rho_\textrm{2K}-\rho_\textrm{min}$, the concentration of Kondo scattering centers $n_\textrm{K}$ could be roughly estimated by $\Delta\rho=2mn_\textrm{K}/[\uppi\hbar ne^2N(\varepsilon_\textrm{F})]$~\cite{Hewson1997book}, for which $n\approx1.76\times10^{21}$~cm$^{-3}$ and $N(\varepsilon_\textrm{F})\approx1.24$~states~eV$^{-1}$~u.c.$^{-1}$ for B8; $m=2.82m_\textrm{e}$ estimated from the cubic phase~\cite{YangF2017MRB}; $\hbar$: the reduced Planck constant. $n_\textrm{K}$ amounts to $3.5\times10^{21}$ cm$^{-3}$, which is rather dense corresponding to 0.2 per u.c.. This result is compatible with the fact that there are sufficient tunneling centers near $n^*$. We note that in SCTO$_{3-\delta}$, $\rho$ shows an upturn below the temperature of polar phase transition at $n<n^*$ and becomes Fermi-liquid-like at $n\approx n^*$~\cite{Rischau2017NP,WangJL2019npj}, in stark contrast with that in BTO$_{3-\delta}$, whose reason  is elusive. We guess the absence of Kondo-like phenomena near $n^*$ is probably due to the much weaker coupling between electrons and lattice in SCTO$_{3-\delta}$~\cite{ChenHH2021NC}.

\vspace{3ex}
\noindent\textbf{DISCUSSION}\\
In the following, the hierarchy of responsivity of different probes to the Kondo effect is discussed. $S$ is expressed by  $S=\alpha/\sigma$,  wherein $\alpha$/$\sigma$ is the thermoelectric/electric conductivity. In light of the semiclassical transport theory~\cite{Behniabook2015},  $\sigma$ and $\alpha$ are expressed by in integrals over energy ($\varepsilon$) (seen in Fig.~\ref{Fig4}h-i):
\vspace{-2 pt}
\begin{equation}
	\begin{split}
	\sigma&=\frac{2e^2}{h} \int d\varepsilon(-\frac{\partial f}{\partial \varepsilon})\Xi(\varepsilon) \\
	\alpha&=-\frac{2ek_\textrm{B}}{h} \int d\varepsilon(-\frac{\partial f}{\partial \varepsilon})\frac{\varepsilon-\mu}{k_\textrm{B}T} \Xi(\varepsilon) \\
		\Xi&=\frac{h}{2}v^2 \tau(\varepsilon) N(\varepsilon)
	\end{split}
	\label{Eq2}
\end{equation}
\noindent wherein $h$ is the Planck constant; $\mu$: the chemical potential; $f$: the Fermi-Dirac distribution; $\Xi$: the transport distribution function; $v$, $\tau$ and $N$: the energy dependent electron velocity, scattering time and DOS.

In Fig.~\ref{Fig4}h, electrons, participating in charge transport, are mostly those at the Fermi level. While, in Fig.~\ref{Fig4}i for thermoelectric transport, the entropy-carrying electrons are those slightly above or below the Fermi level. If $\Xi(\varepsilon)$ slightly evolves (in most cases), the contribution to $\alpha$  from both sides of the Fermi level mostly cancels each other out. Once the dipolar-Kondo effect produces an Anderson-Friedel resonance on DOS, asymmetric with respect to the Fermi level~\cite{Hewson1997book}, a large amount of entropy taken by this resonance cannot be balanced and leads to a manifold enhancement of $S$, as in Fig.~\ref{TEP}b. In Fig.~\ref{Fig4}h, this resonance only donates a small part of charge-carrying electrons, thus its impact on $\rho$ is modest as in Fig.~\ref{Fig4}d. %which is asymmetric with respect to the Fermi level in absence of particle-hole symmetry

This picture also gives a good account of sign mismatch between $S$ and $R_\textrm{H}$. The sign of $R_\textrm{H}$ is set by the average of local band curvature at the Fermi surface. While, the sign of $S$ can be either positive or negative, profoundly influenced by the resonance position with respect to the Fermi level.  $\gamma$ is yielded by extrapolating $C_\textrm{P}/T$ to 0 K, for which the entropy-carrying electrons are at the Fermi level. Thus, the entropy taken by the resonance hardly contributed to $\gamma$, in line with the rather smooth evolution of $\gamma$ in Fig.~\ref{TEP}d. Overall, $S$ is the most sensitive probe of Kondo physics. This result may give a clue to enhanced $S$ at the P-NP transition in Mo$_{1-x}$Nb$_{x}$Te$_2$~\cite{Ishiwata2016SA}.  Here, we attain the second outcome of the paper.

In the end, let us discuss the downturn on $\rho$ at $T\rightarrow 0$~K for specimens around $n^*$. The answer is still open. We present below three possibilities. First of all, it might be associated with magnetism induced by $V_\textrm{O}$~\cite{Yang2010SCPMA}. However, the magnetic susceptibility does not show any observable transition at similar $T$ in Supplementary Fig.~10. Second, recent DFT calculations predicted SC with $T_\textrm{c}\approx2$~K near $n^*$ in BTO~\cite{ChenHH2021NC}. While, in Supplementary Fig.~8, $\rho$ remains finite at $T=50$ mK. One may guess SC is destroyed by strong impurity scattering in present system. Third, given the \textcolor{black}{potential} dipolar-RKKY interaction between dense tunneling centers, one may imagine a Kondo coherence below a characteristic temperature, reminiscent of Kondo lattice in heavy Fermions~\cite{Hewson1997book}. If this is true, one would have observed features on other coefficients.

In summary, our experiments demonstrate a doping threshold for P-NP transitions, regardless of physical details in distinct polar metals. Such a doping is interpreted by a dipolar-RKKY interaction. Near $n^*$, the anomalous enhancement on $S$ and the upturn on $\rho$ imply the possibility of the dipolar-Kondo scenario. A comprehensive theory with respect to the dipolar-RKKY and dipolar-Kondo begs for future investigation.

\vspace{3ex}
\noindent\textbf{METHODS}\\
The study was performed on commercial BTO single crystals. Oxygen deficient BTO$_{3-\delta}$ was obtained by annealing stoichiometric samples, which were enclosed by high purity Ti powders within a tantalum crucible and then sealed in a quartz tube. For various electron concentrations, the tube was heated in an oven for two hours by varying $T$ from 750~\textcelsius ~to 1100 ~\textcelsius. Ti (5nm)/Au (60nm) electrodes were deposited by electron beam evaporation to achieve Ohmic contact.

Variable $T$ XRD measurements were performed in a Bruker D8 X-ray diffractometer equipped with a closed-cycle He cryostat system. Dielectric constants were measured by Hioki IM3536 LCR meter. Transport and heat capacity measurements were carried out in Quantum Design physical property measurement system (PPMS-16T). Thermoelectric power was measured by using a steady-state configuration with one heater and two thermocouples. Resistivity at sub-kelvin was measured in an Oxford dilution refrigerator.  DC magnetization was measured in a Quantum Design magnetic property measurement system (MPMS3).

\vspace{4ex}
\noindent\textbf{DATA AVAILABILITY}\\
The data that support the findings of this study are available from the corresponding author upon reasonable request.

\vspace{3ex}
\noindent\textbf{ACKNOWLEDGMENTS}\\
We are grateful to Kamran Behnia, Jinke Bao and Shi Liu for the helpful discussion. This research was supported by Zhejiang Provincial Natural Science Foundation of China for Distinguished Young Scholars under Grant No. LR23A040001, National Natural Science Foundation of China via Project 11904294, Zhejiang Provincial Natural Science Foundation of China under Grant No. LQ23A040009 and Postdoctoral Science Foundation of Zhejiang Province under Grant No. ZJ2022094. We thank the support provided by Dr. Xiaohe Miao and Dr. Chao Zhang from Instrumentation and Service Center for Physical Sciences at Westlake University.

\vspace{3ex}
\noindent\textbf{COMPETING INTERESTS}\\ 
The authors declare no competing interests.\\

\vspace{3ex}
\noindent\textbf{AUTHOR CONTRIBUTIONS}\\
X.Y. and W.H. prepared the samples. X.Y. did the XRD and transport measurements assisted by J.W., T.W. and Z.L.. Z.X. did the dielectric measurements. W.H., X.Y. and X.L. prepared the figures.  X.L. led the project and wrote the paper.  All authors contributed to the discussion.

\vspace{2ex}
\noindent\textbf{ADDITIONAL INFORMATION}\\

\noindent\textbf{Supplementary information} accompanies the paper on the npj Quantum Materials website (doi:...).

%\bibliographystyle{naturemag}
%\bibliography{BTO}
\vspace{3ex}
\noindent\textbf{REFERENCES}\\

\end{document}